\documentclass[runningheads]{llncs}
\usepackage{graphicx}
\usepackage{float}
\begin{document}
\title{Examining Different Research Communities: Authorship Network \thanks{Thanks to my advisor Dr. Bojan Cukic}}
%
%
\author{Shrabani Ghosh\inst{1}\orcidID{0000-0002-6084-4964} 
}
\authorrunning{Shrabani Ghosh}
%
\institute{University of North Carolina at Charlotte, NC 28223, USA \and
\email{sghosh15@uncc.edu}}
\maketitle              
\begin{abstract}

Google Scholar is one of the top search engines to access research articles across multiple disciplines for scholarly literature. Google Scholar's advanced search option gives the privilege to extract articles based on phrases, publisher name, author name, time duration, etc. In this work, we collected Google Scholar data (2000-2021) for two different research domains in computer science: Data Mining and Software Engineering. The scholar database resources are powerful for network analysis, data mining, and identifying links between authors via co-authorship network. We examined co-authorship networks for each domain and studied their network structure. Extensive experiments are performed to analyze publication trends and identify influential authors and affiliated organizations for each domain. The network analysis shows that the network's features are distinct from one another and exhibit small communities within the influential authors of a particular domain.   

\end{abstract}
\section{Introduction}

Social network analysis has become a popular technique in many fields such as social media networks, web networks, biological networks, academic social networks etc. The network analysis gives a broader picture of connections that has been maintained within a community over a long period of time. As academic papers are increasing exponentially, each academic area is segmented and specialized. To better understand the scientific community in general, it is important to study and model the factors of research communities, the growth and evaluation of a particular research domain. Therefore, modeling these factors efficiently is important to better understand the working culture of the scientific community. In that case, authorship network is very useful for analysis across different scientific domains. We are interested to look at the authorship network of research domains under computer science stream in recent era. The pattern and culture are quite a bit different in different research domains under same stream. Most of the authorship networks analysis finds communities and structures of sub-communities.  

Many co-authorship networks have been studied to investigate collaboration among users \cite{cunningham1997authorship,egghe2000methods,newman2001scientific}. Most of them used databases, for example Aminer, DBLP digital libraries to generate network. 
DBLP provides comprehensive access to conference proceedings and allows for filtering of publications by conference stream, offering similar functionality to parsing results from Google Scholar.
Co-authorship network can describe publication rate and scientific productivity \cite{costa2014dynamics,melin2000pragmatism}. Previous studies have shown that, there is a strong positive relation between co-authorship network and  scientific productivity \cite{leydesdorff2013international,glanzel2004analysing}. Researchers from diverse geographical locations \cite{goldfinch2003science,nemeth2005creative} and multi-disciplinary backgrounds collaborate together which positively impacts scientific productivity. Social network studies have furthered the understanding of the relationship between co-authorship and productivity. Studies assessing the relationship between productivity and the position of authors in the co-author network have found that authors who publish with many different co-authors bridge communication and exhibits higher rates of publication records. 
 
In this paper, we use data collected from Google Scholar on data mining and software engineering research domain. Using this dataset, our proposed work will follow the below research contributions:
\begin{itemize}
    \item Investigate pattern of publication over past 20 years for different research domains.
    \item Identify the most influential authors based on published articles.
    \item Identify the most frequently affiliated organization appeared  within a domain.
    \item Analyze and compare network features of two distinct research areas.
    
\end{itemize}

\section{Related Work}
Co-authorship networks have been widely used in the past to explore scientific collaboration among authors. An authorship network is an academic social network where individuals are connected based on their collaboration through research articles. In previous studies, co-authorship networks have been analyzed to explore different statistical characteristics. It has been extensively analyzed at network level and individual node level~\cite{newman2001structure} and behavioral patterns in scientific collaborations \cite{liu2005co}. Similar to authorship network, citation network also have been studied much. In citation network, nodes connection are created from paper citations, where papers are nodes and links generate when one paper is cited by another paper. Usually citation network shows the relation among papers. On the other hand, Co-authorship networks illustrate the relationships among researchers, typically reflecting collaborative bonds formed through joint publications. These networks are dynamic, evolving as researchers collaborate within or across different research domains over time. A single researcher may collaborate in multiple research areas, thereby participating in different co-authorship networks. While these networks may overlap due to the presence of common researchers, they can still represent distinct collaborative groups based on the specific research topics or domains. Many research dimensions like finding rising stars \cite{li2009searching}, name disambiguation \cite{huang2014institution}, domain experts \cite{daud2010temporal} have been studied in academic social network.

In this paper, we analyze and compare the publication pattern, features of domain experts and collaboration culture of two research domains in computer science. The number of published papers per year shows how these two fields have emerged over time. From the google scholar profile, it is easy to capture authors information and affiliations to understand the most involved organizations/institutions in a domain. Also, the network analysis shows relation among authors and degree distribution of authors. 
\section{Data Collection}
The dataset used in this work is collected from Google Scholar. The data is collected for two different research domains: Data Mining and Software Engineering. The focus on software engineering and data mining in this analysis was chosen to offer a comparative perspective between two distinct domains: one well-established and foundational (software engineering) and one that has experienced significant growth in recent years (data mining). Software engineering, being a more mature field, allows us to examine long-term trends and shifts in publication practices, while data mining, a relatively younger and increasingly popular research area, provides insights into how emerging fields evolve and attract academic and industrial interest.


For each research domain, three recognized and popular conferences have been chosen to extract using advanced search option. The three selected conferences for Data Mining are, 1. IEEE International Conference on Machine Learning and Applications (ICMLA)
2. IEEE International Conference on Data Mining (ICDM)
3. ACM International Conference on Knowledge Discovery and Data Mining (SIGKDD). On the other hand, for Software Engineering the selected conferences are: 1. International Conference on Software Engineering (ICSE)
2. ACM International Symposium on Foundations of Software Engineering (SIGSOFT) 
3. IEEE/ACM International Conference on Automated Software Engineering (ASE).
For each conference, we have extracted 1000 published papers within time-frame 2000 to 2021. We do think that the sample 
In total, we have collected 3000 published papers for each area. For dataset interested researchers can contact the author via email\footnote{sghosh15@uncc.edu}. The source code used in this study will be made available in the following directory\footnote{\url{https://github.com/srbnghosh99/Examining-Different-Research-Communities-Authorship-Network}}

A published paper typically involves one or more authors. We have listed authors' information from the papers. We can identify 2788 authors in Software Engineering conferences and 4245 authors in Data Mining conferences. We have collected the author's information: full name, citation, h-index, and affiliation from their Google Scholar profile. We have used this information for further analysis. 

\section{Methods and Results}

The abstract summarizes the idea of the article. It describes the purpose and content of the research paper adequately in fewer words and definitely the first read. To identify the most frequently used words in the title of the articles of a domain, we have created a word cloud based on frequently used listed words. Figure 1 shows simple weighted word clouds of data mining and software engineering based on research article titles. Following these words, data mining research communities seem interested in using deep learning approaches for prediction-based tasks for example classification. On the other hand, software engineering communities are more focused on systems, security, and testing for analysis. 
\begin{figure}
\centering
\begin{minipage}{.5\textwidth}
  \centering
  \includegraphics[width=6cm ]{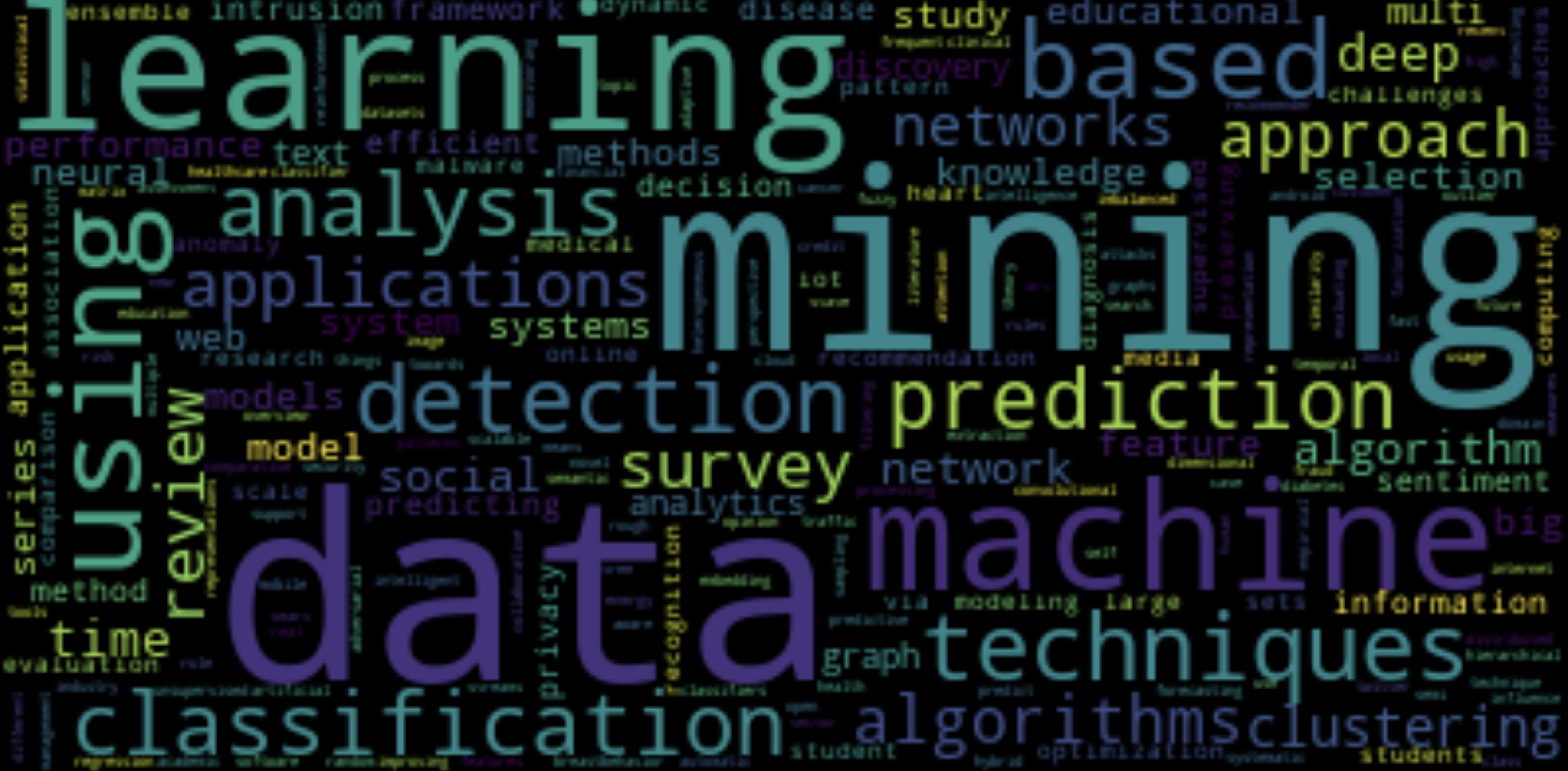}

\end{minipage}%
\begin{minipage}{.5\textwidth}
  \centering
  \includegraphics[width= 6cm]{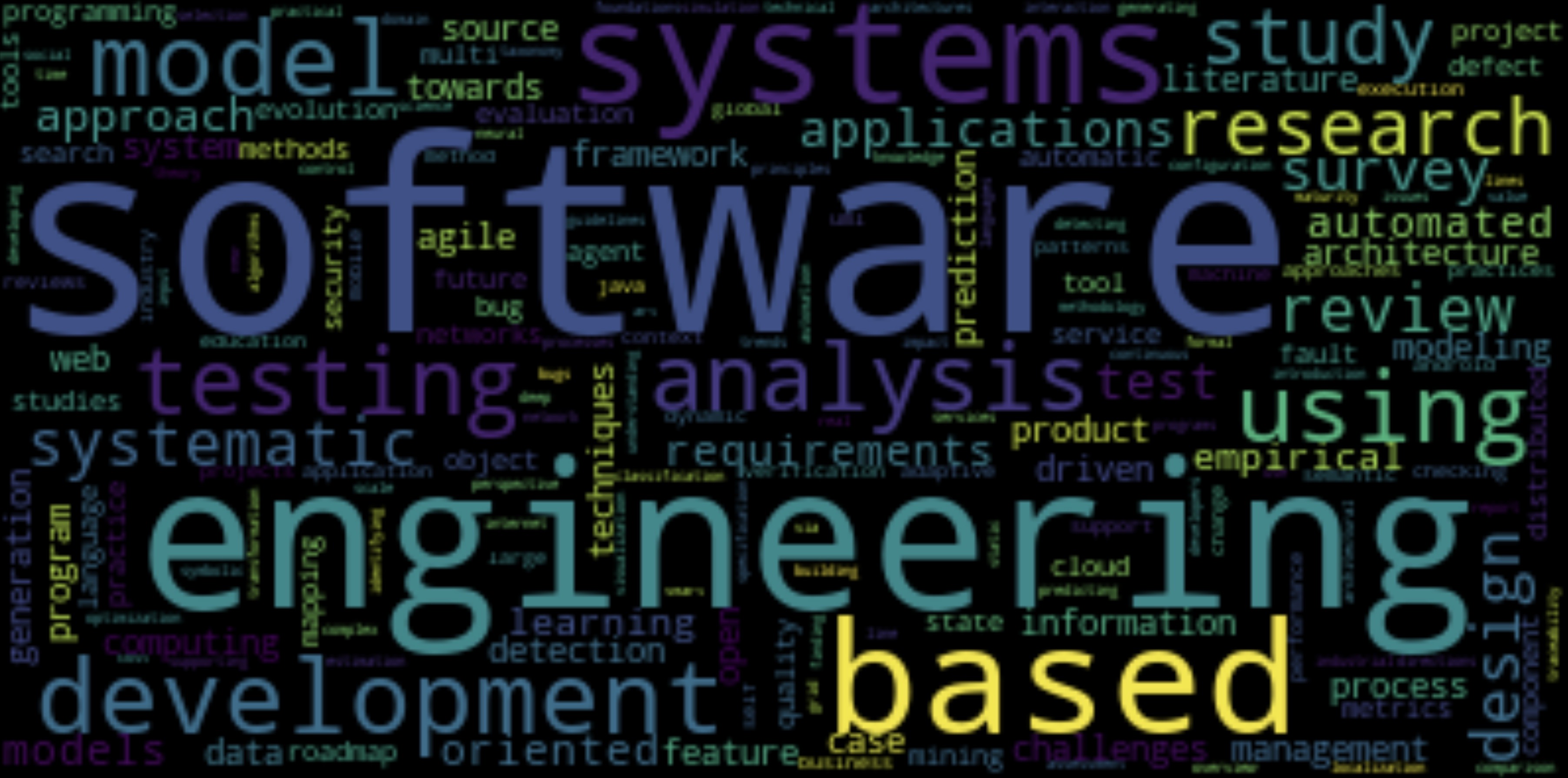}

\end{minipage}
\caption{Weighted wordcloud from the original title. (a) Data Mining (left), (b) Software Engineering (right)}
\end{figure}
\subsection{Paper Publication Pattern}
In order to understand the publication record of these two domains, we accumulated the number of published papers each year within the time frame. The trend shows how the two domains differ in terms of paper publishing over time. Figure 2 highlights that data mining publication records have steadily increased, reflecting growing attention and interest from researchers in this domain. On the other hand, the number of publications in software engineering has declined over time. This contrast demonstrates how one research area is emerging more rapidly than the other.

The data shows that publication in data mining conferences significantly increased from 2012, reaching its highest peak with 312 published papers in 2018. In contrast, software engineering conferences peaked earlier, with 238 papers in 2005, followed by a steady decline, reaching the lowest number of publications in 2020 and 2021.

While these trends are informative, it is important to consider alternative explanations for these differences. One factor could be changing acceptance rates, which may have remained constant or varied differently between the two fields. Also, higher number of researchers in data mining than software engineering could be another reason of decreased number of publications. It is also possible that researchers are getting more interested in data mining over time. Additionally, the emergence of competing conferences in software engineering might have split the pool of submissions, contributing to the observed decline. Conversely, the increasing focus on data mining, driven by the rise of big data and machine learning applications, may have spurred more publications in that field. Although this study does not examine these factors in detail, they offer plausible explanations for the publication trends observed in Figure 2.
\begin{figure}
\centering
\begin{minipage}{.5\textwidth}
  \centering
  \includegraphics[width=6cm ]{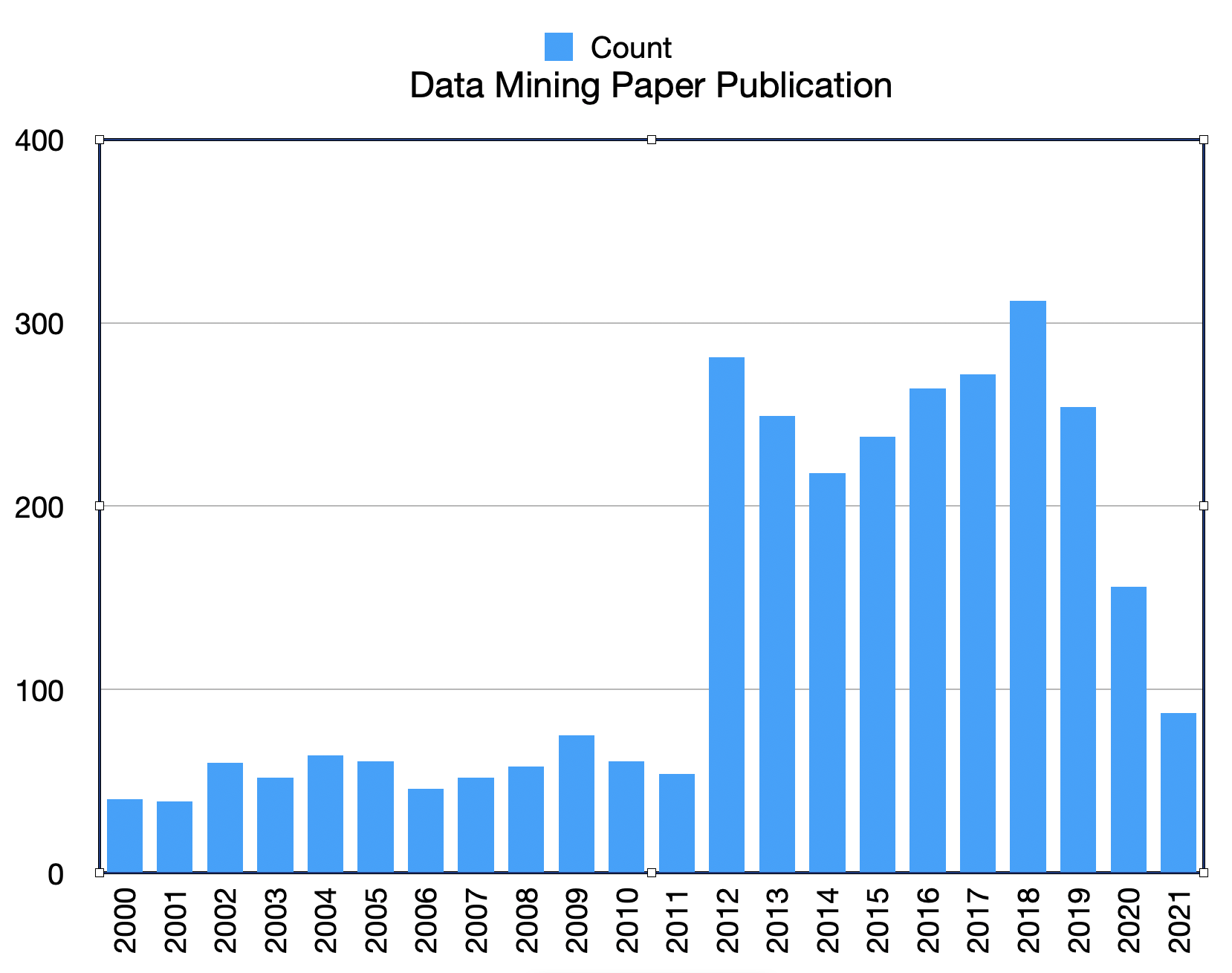}
\end{minipage}%
\begin{minipage}{.5\textwidth}
  \centering
  \includegraphics[width= 6cm]{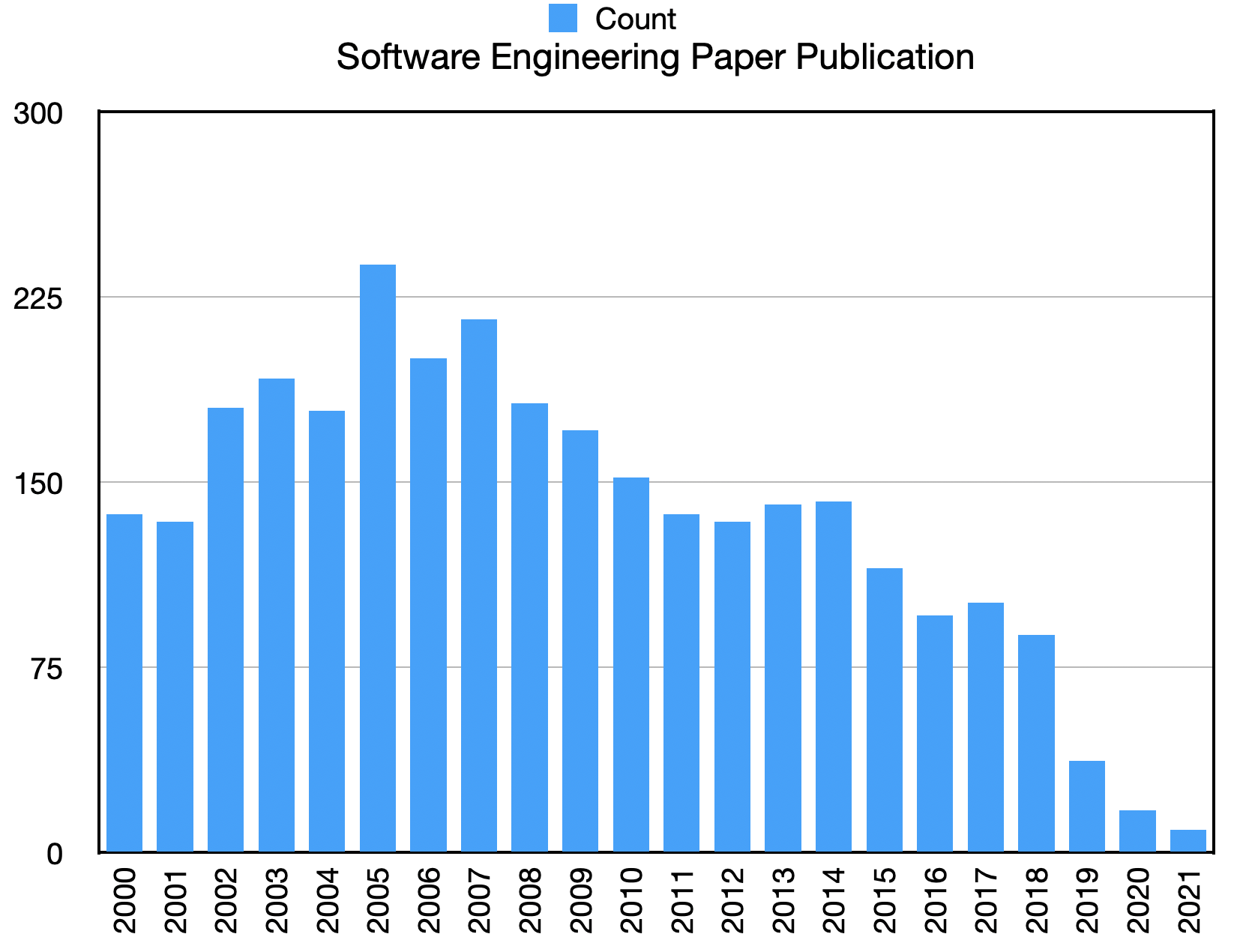}
\end{minipage}
\caption{Number of publications per year: The left image shows Data Mining and right image shows Software Engineering}
\end{figure}

\subsection{Influential Authors}
Authors tend to publish papers at the same conferences consistently, allowing us to identify the influential authors based on the number of papers published within their respective research domains. In the scholar data, each author is assigned a unique identifier. Our analysis is conducted based on these unique IDs, allowing us to subsequently map them back to their corresponding original names.
In this study, we define influence by how frequently an author appears as either the first author or a co-author. Using this metric, we have identified the top 10 influential authors based on their publication records in our dataset. From Figure 3, the top 5 influential authors in data mining are Jiawei Han, Huan Liu, Eamonn Keogh, Philip Yu, and Ryan Baker, with Jiawei Han appearing 32 times and Huan Liu 30 times between 2000 and 2021. In software engineering, Barbara Kitchenham, Thomas Zimmermann, Mark Harman, Gail Murphy, and Krysztof Czarnecki stand out, with Barbara Kitchenham appearing 35 times and Thomas Zimmermann 26 times in the same period. The influence of these authors reflects not only their publication frequency but also their significant contributions to their fields. For instance, Jiawei Han’s foundational work in data mining, particularly in developing algorithms for large-scale datasets, and Barbara Kitchenham’s systematic approaches to evidence-based software engineering, have made them central figures in their domains. While it is important to recognize these key contributors, the analysis focuses on their impact and contributions rather than merely listing their names. This ensures that the recognition of influential researchers is grounded in their significant and consistent output over time

\begin{figure}[tbh]
\centering
\begin{minipage}{0.5\textwidth}
  \centering
  \includegraphics[width=\textwidth ]{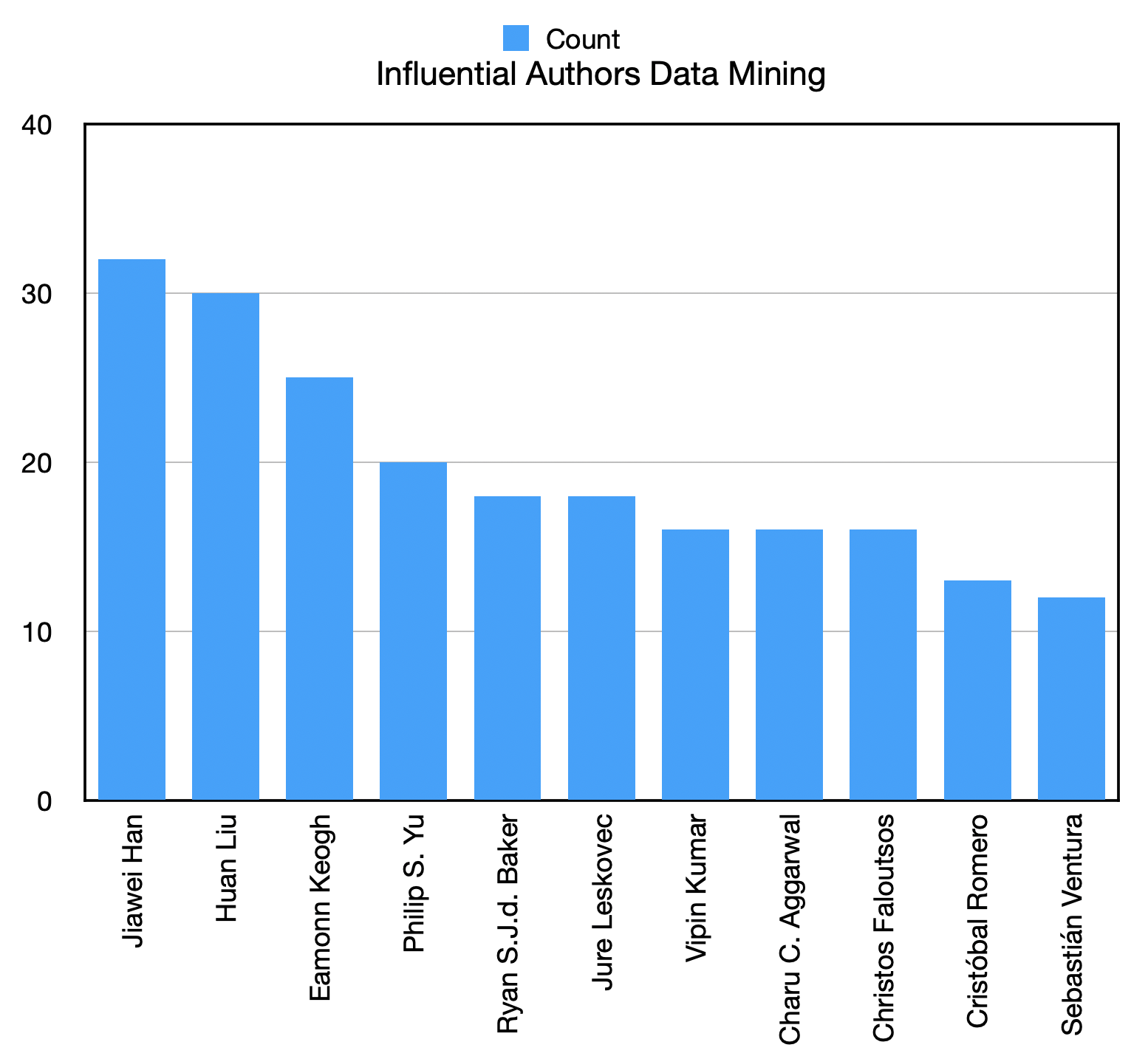}

\end{minipage}%
\begin{minipage}{0.5\textwidth}
  \centering
  \includegraphics[width=\textwidth]{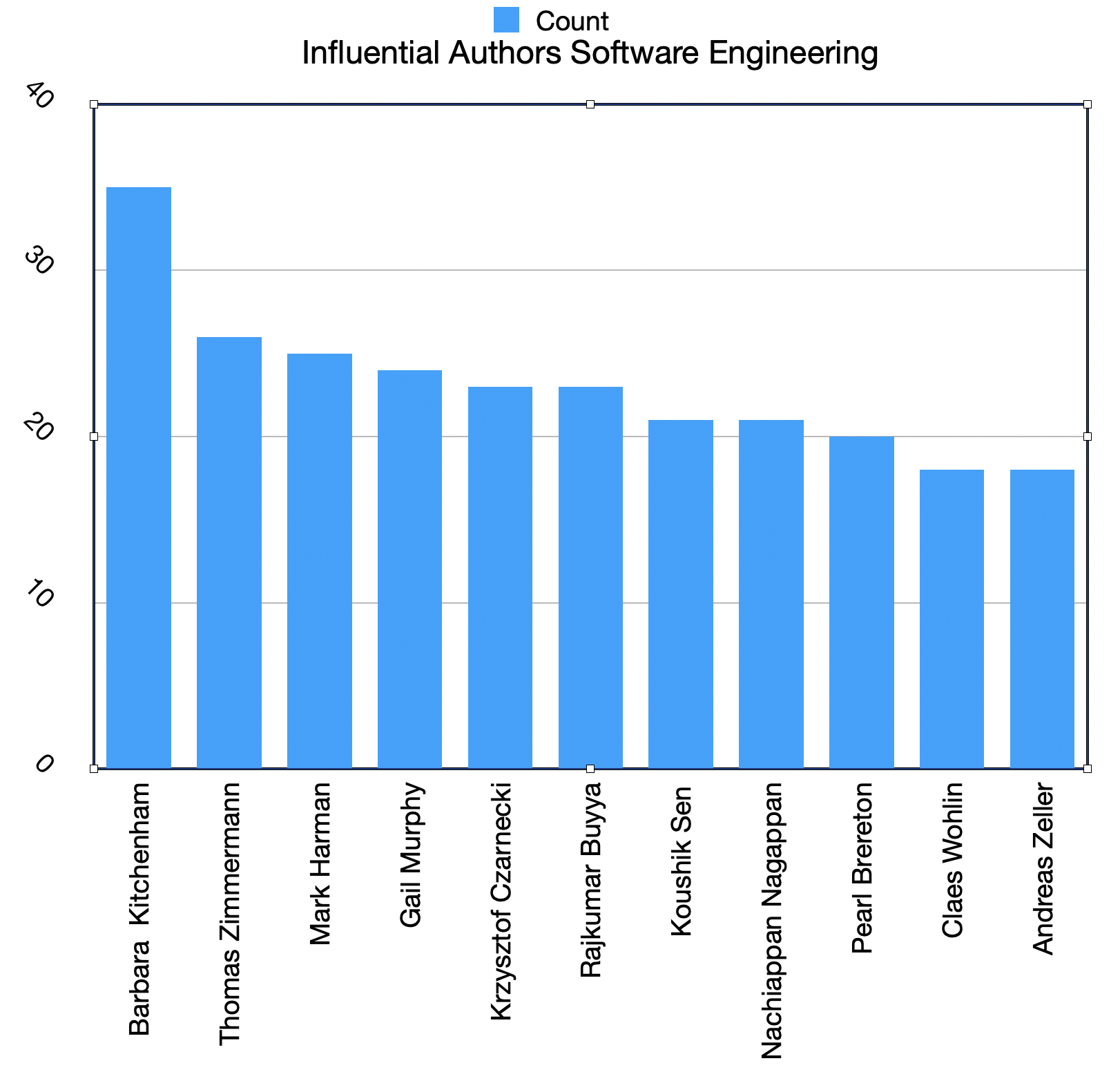}

\end{minipage}
\caption{The top 10 influential authors: The left image shows Data Mining and right image shows Software Engineering}
\end{figure}



\subsection{Affiliated Organizations}
We also looked at the authors Google Scholar profiles that are classified as valid profiles. The valid profiles usually have full name, research interest, citations, h-index, and affiliated organization information. We have captured the current or available affiliations of the authors as provided by Google Scholar. However, we do not have access to the authors' affiliations at the time of publication
This limitation may introduce distortions, as authors often change institutions over time. For example, an author like Barabási may have been affiliated with different organizations during the publication of various papers, yet only the most recent or selected affiliation might be displayed in databases like Google Scholar. 

From the most current affiliation collected from Google Scholar, we have come up with the top most frequently appeared affiliated organizations of the authors for each research domain. In data mining domain, we can see that Google has been the topmost affiliated company to collaborated with, then University of Minnesota, Microsoft, University of Illinois Urbana-Champaige, University of North California, Riverside so and so forth. In software engineering, Microsoft has been the topmost affiliated company, and then Keele University, Carnegie Mellon University, Facebook, Waterloo University so and so forth (Figure 4).

These affiliations provide valuable insights into the institutions that have a significant influence on research in these fields. For instance, Google, Microsoft and Facebook have prominent roles in both domains reflect their ongoing investment in research and development, as well as their collaborations with leading academic institutions. The presence of Carnegie Mellon University further highlights the contribution of academic research to both data mining and software engineering. This analysis underscores how industry-academia partnerships shape research trends and suggests that institutions with strong affiliations in these domains are likely to remain influential in advancing research and innovation in the future.

\begin{figure}[tbh]
\centering
\begin{minipage}{0.5\textwidth}
  \centering
  \includegraphics[width=\textwidth ]{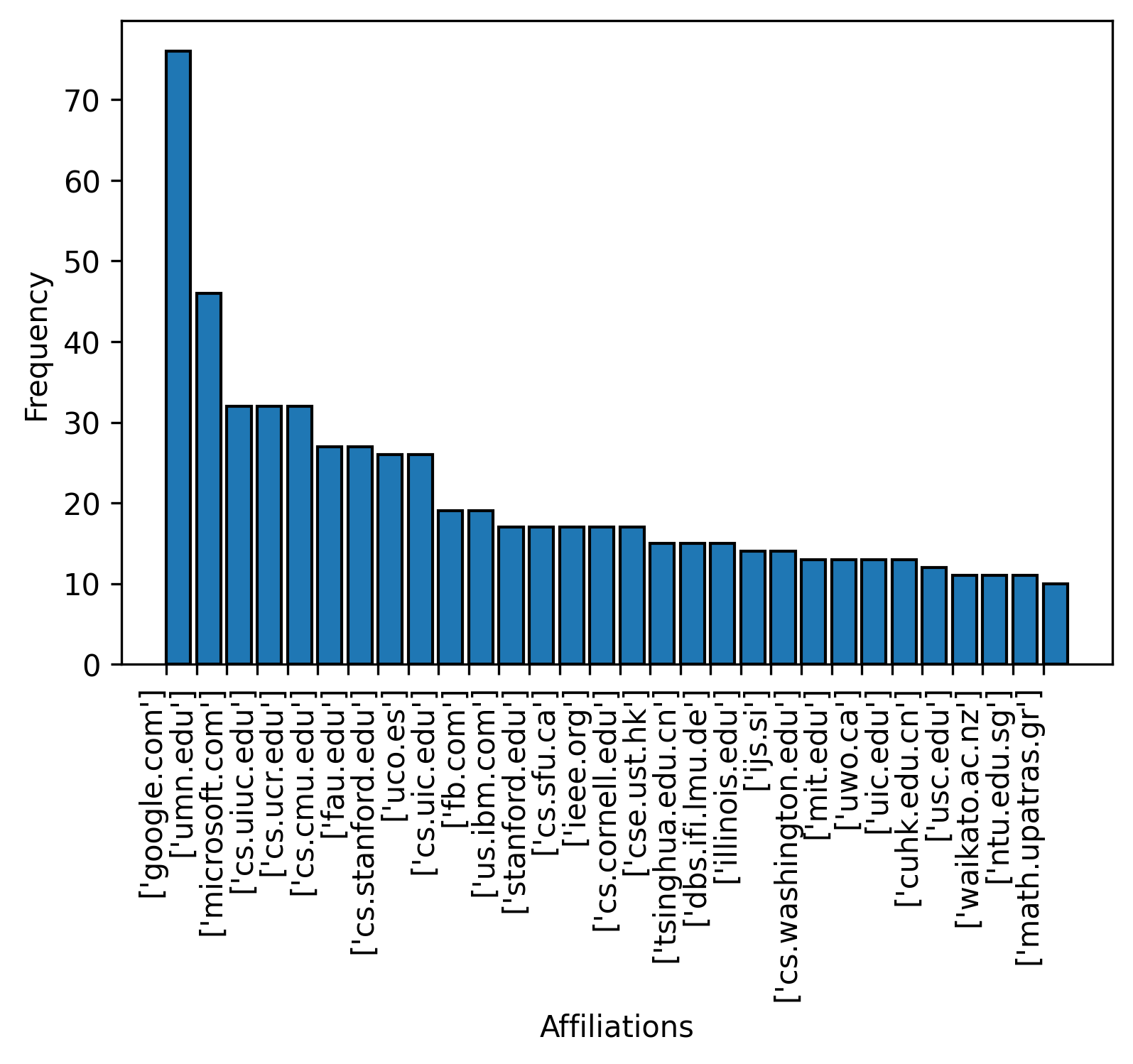}

\end{minipage}%
\begin{minipage}{0.5\textwidth}
  \centering
  \includegraphics[width=\textwidth]{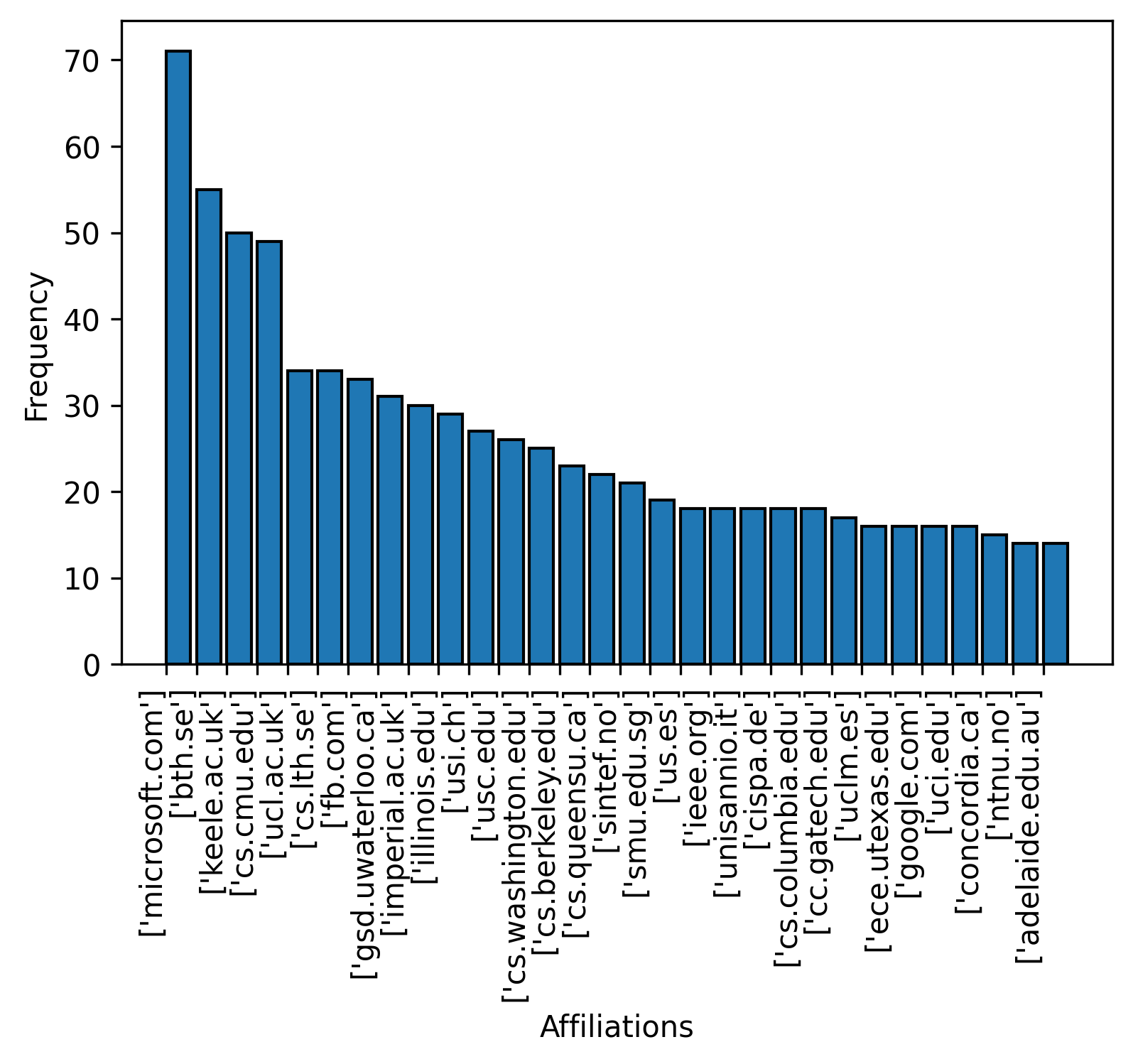}

\end{minipage}
\caption{Number of publications per year: The left image shows Data Mining and right image shows Software Engineering}
\end{figure}

\section{Network Analysis}
In this section, we perform a co-author network analysis to examine the relationships between the authors within domains. We have generated undirected network for both domains. 
Data mining co-author network has 4245 nodes and 4363 edges. 
There are 1523 connected components in the network and the giant connected component has 571 nodes involved within it. Figure 5 (a) shows the largest connected component on the top and degree distribution for all the nodes. The distribution shows that more than 1200 nodes have degree value of 2 and few nodes have degree value above 10. We found Huan Liu and Jiawei Han have highest degree value of 44 \& 39 correspondingly.

On the other hand, the software engineering co-author net has 2788 nodes and 2870 edges. Many papers have only one author. Out of the whole network, there are 898 connected components. The giant connected component of the network consists of 708 nodes. The degree histogram (Figure 5 (b)) of the whole network shows that more than 800 nodes have degree 2 and few nodes have degree 10. In this case, Mark Harman and David Lo have degree value of 21 and 20 correspondingly.

Compared to software engineering, data mining authors' highest degree value 40 is twice that of software engineering authors' highest degree value 20.

\begin{figure}[tbh]
\centering
\begin{minipage}{0.5\textwidth}
  \centering
  \includegraphics[width=\textwidth ]{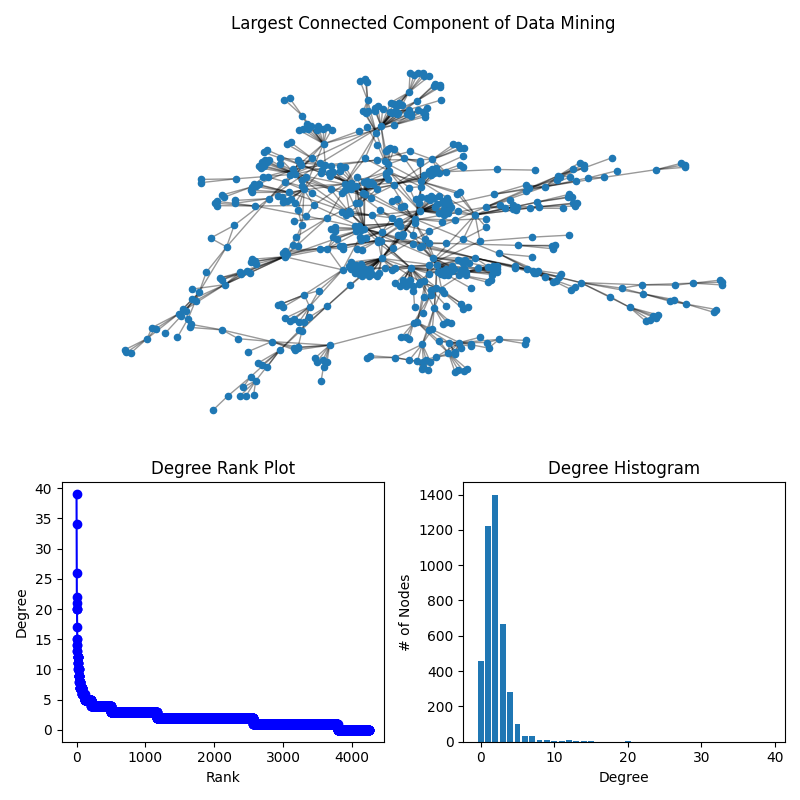}

\end{minipage}%
\begin{minipage}{0.5\textwidth}
  \centering
  \includegraphics[width=\textwidth]{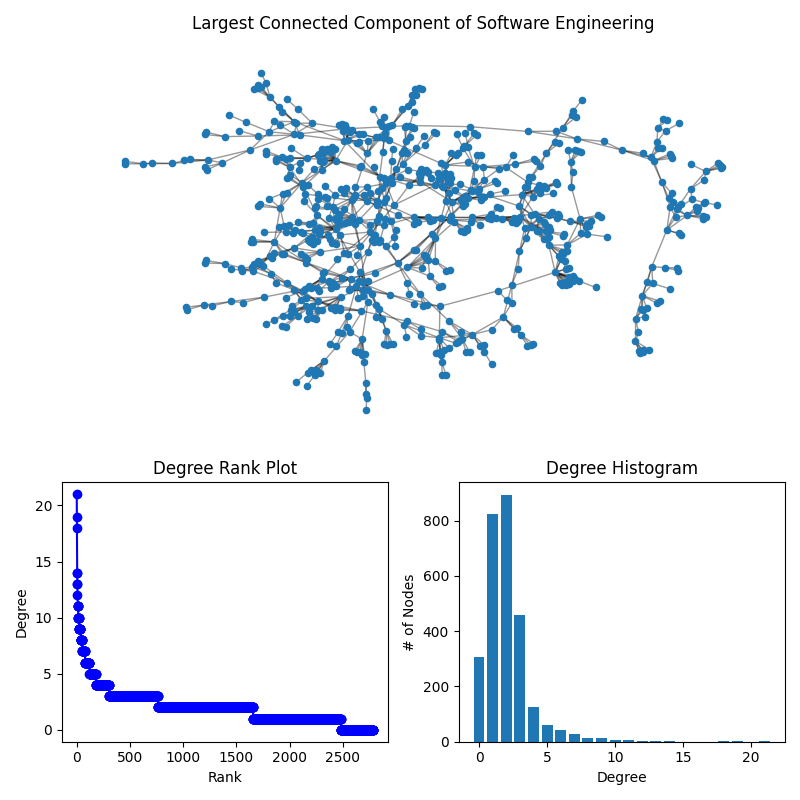}

\end{minipage}
\caption{Degree Distribution: (a) The left image shows Data Mining and (b) right image shows Software Engineering}
\end{figure}


\subsection{Collaboration Teams}
In today's research system, team collaboration is the most essential part. In a research collaboration team, collaborators interchange their ideas, thoughts, datasets, problem-solving approaches, writing, etc. Some of the effective teams work together many times. It is important to point out strong tied collaboration teams in the co-author network to understand collaboration culture in a research domain.  

From the network analysis, we have brought out the paired collaborators who co-authored together several times. We found that in software engineering, Barbara Kitchenham has collaborated with Pearl Brereton highest 13 times and with David Budgen 12 times. Also, "Per Runeson and Martin Höst", "Nachiappan Nagappan and Thomas Ball", "Andrea De Lucia and Rocco Oliveto" each paired collaborators worked together 8 times within the time-frame for the conferences mentioned early. 

On the other hand, in data mining "Cristóbal Romero and Sebastián Ventura" collaborated together highest 12 times, "Jiliang Tang and Huan Liu" 7 times. "George Karypis and Joseph Konstan", "Eamonn Keogh and Stefano Lonardi", 
"Igor Santos and Xabier Ugarte-Pedrero" each paired collaborator worked together 5 times. 
\subsection{Connected Authors}
At a high level, connected components exhibit several key characteristics, often representing small communities within the network. Our focus here is on closely examining the largest connected components from each domain. The size and color of the nodes are determined by their degree, while the weight of the edges reflects the number of times two authors have collaborated.

Figures 6 and 7 allow for a comparison of the connected components across these two domains. In software engineering, authors tend to be sparsely connected, with fewer instances of tightly knit collaborations. Conversely, in data mining, the nodes are more densely interconnected, indicating that several authors have frequently worked together. This is particularly evident in the purple-colored edges, which represent high levels of collaboration.

\begin{figure}[!h]

  \centering
  \includegraphics[width=10cm]{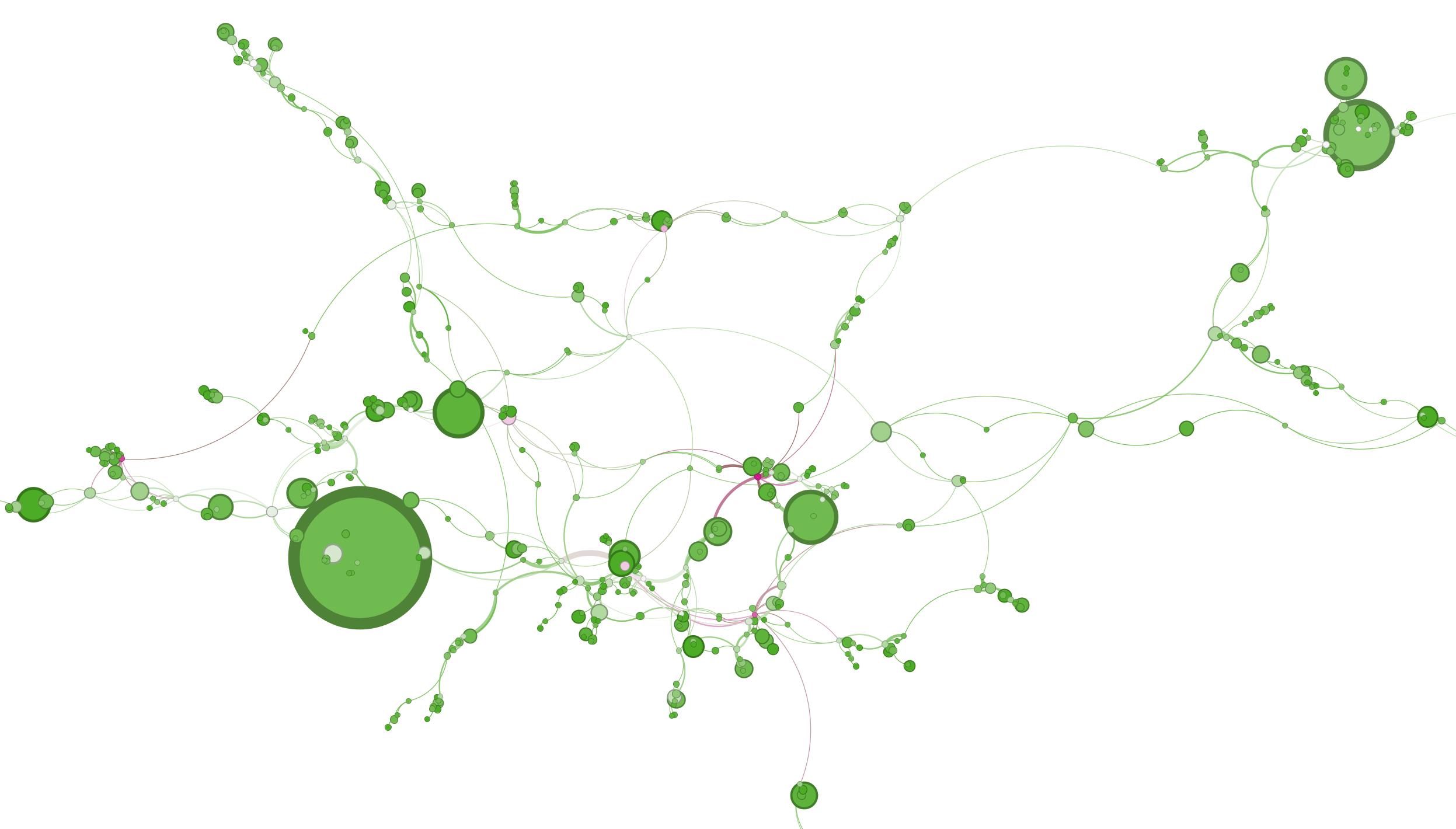}
  \caption{Software Engineering Largest Connected Component}
\end{figure}

\begin{figure}[!h]

  \centering
  \includegraphics[width=10cm]{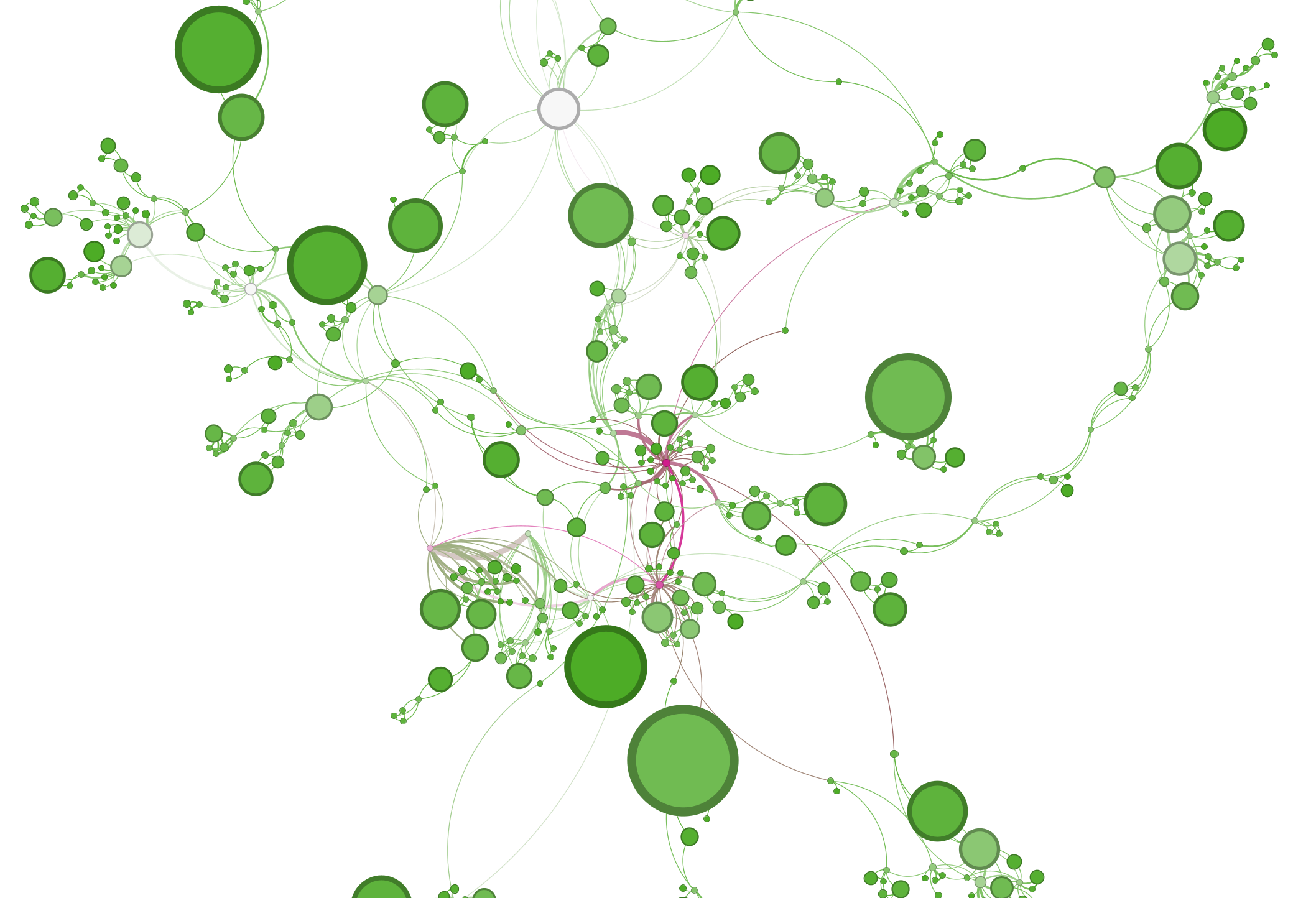}
  \caption{Data Mining Largest Connected Component}
\end{figure}

 

\section{Conclusion}
We studied the characteristics of two different research communities: data mining and software engineering authorship network from google scholar. Analyzing network structure, collaboration and author profiles of these two communities allowed us to discover interesting features. We have observed that data mining research domain has flourished more than software engineering over past 10 years. Though we have seen that the influential authors have similar number of publication record in both research domain, however, in data mining have more researchers compared to software engineering. We have seen that from the network analysis that data mining authors have more knit group of collaboration than software engineering authors. The overall collaboration culture and number of authors helped to flourish data mining domain. Furthermore, the giant tech company for example Google collaborators seemed to be more interested in data mining domain and also Microsoft collaborators have higher publication record in data mining than software engineering tracks.  

Although our findings are based solely on data from selective conferences, and we do not intend to generalize these results broadly, we believe they offer a meaningful perspective on the trends and developments within the particular domains covered by these conferences.
We wanted to present findings and demonstrate it through our dataset. The dataset used was carefully selected to ensure that it is representative of the field. Nevertheless, we acknowledge that future work could benefit from exploring networks constructed from different or broader samples to evaluate whether the observed patterns hold. This process can be extended to many directions in future. Authorship networks can be useful for predicting future collaboration teams by analyzing historical collaboration patterns among authors. By examining the connections between authors in the network, we can identify individuals who have previously collaborated, which often indicates shared research interests and complementary expertise. Additionally, network analysis techniques can reveal clusters of authors who are frequently connected, suggesting potential teams for future collaborations. This approach aligns with social network theory, which posits that individuals are more likely to collaborate with those within their existing networks.
Authorship network can be useful to do further network analysis. For example, centrality measures to rank authors not just by publication count, but by their importance in the co-authorship network. Also, as a dynamic network we can study study the evolution of the network for each domain over time: if there are any emerging trends or shifts in key players.  
We are highly interested to explore citations network of each domain and analyze citation authors and articles network. As a future direction, we propose to develop a comprehensive analysis pipeline that can be adapted for studying various research areas. This pipeline will involve customizing the selection of conferences or journals and applying similar methodologies to analyze the co-authorship and citation networks in different fields.


In addition to the primary findings of this study, we recognize that understanding authorship networks can illuminate broader issues, including gender and racial biases in research collaboration. Our analysis highlights the collaborative patterns among authors, which may reflect underlying biases in the academic community. Future work will aim to investigate how these biases manifest in co-authorship networks, potentially impacting the diversity of research teams and the representation of marginalized voices in scientific discourse. By exploring these aspects, we can gain a more comprehensive understanding of the factors that shape collaborative practices and promote equity in research.

\bibliographystyle{plain}
\bibliography{ref.bib}








\end{document}